\newcommand{\largesize}{}
\documentclass[12pt]{article}

\usepackage{scicite}
\usepackage{times}
\usepackage{amsmath}
\usepackage{graphicx}
\usepackage{url}
\graphicspath{{./figures}}

\newenvironment{sciabstract}{%
\begin{quote} \bf}
{\end{quote}}

\title{\largesize{In situ observations of large amplitude Alfv\'en waves heating and accelerating the solar wind}}
\author
{Yeimy J. Rivera$^{1\ast\dagger}$, Samuel T. Badman$^{1\dagger}$, Michael L. Stevens$^{1}$,
Jaye L. Verniero$^{2}$, \\ Julia E. Stawarz$^{3}$, Chen Shi$^{4}$,
Jim M. Raines$^{5}$, Kristoff W. Paulson$^{1}$, \\ Christopher J. Owen$^{6}$,  Tatiana Niembro$^{1}$, Philippe Louarn$^{7}$, Stefano A. Livi$^{8,5}$, \\ Susan T. Lepri$^{5}$, Justin C. Kasper$^{9,5}$, Timothy S. Horbury$^{10}$, \\ Jasper S. Halekas$^{11}$, Ryan M. Dewey$^{5}$, Rossana De Marco$^{12}$, Stuart D. Bale$^{13}$
\\
\footnotesize{$^{1}$Center for Astrophysics, Harvard \& Smithsonian,}\\
\footnotesize{60 Garden Street, Cambridge, MA 02138, USA}\\
\footnotesize{$^{2}$Code 672, NASA, Goddard Space Flight Center, Greenbelt, MD 20771, USA}\\
\footnotesize{$^{3}$Department of Mathematics, Physics, and Electrical Engineering,} \\
\footnotesize{Northumbria University, Newcastle upon Tyne, NE1 8ST, UK}\\
\footnotesize{$^{4}$Department of Earth, Planetary, and Space Sciences,} \\ 
\footnotesize{University of California, Los Angeles, CA 90095, USA}\\
\footnotesize{$^{5}$Department of Climate and Space Sciences and Engineering, University of Michigan,}\\
\footnotesize{2455 Hayward Street, Ann Arbor, MI 48109, USA}\\
\footnotesize{$^{6}$University College London/Mullard Space Science Laboratory,} \\
\footnotesize{Holmbury St Mary, Dorking, Surrey, RH5 6NT, United Kingdom}\\
\footnotesize{$^{7}$Institut de Recherche en Astrophysique et Planétologie/Center for Scientific Research,} \\ 
\footnotesize{University of Toulouse, Toulouse, France}\\
\footnotesize{$^{8}$Southwest Research Institute, 6220 Culebra Rd, San Antonio, TX 78228, USA}\\
\footnotesize{$^{9}$BWX Technologies Inc, Washington, DC 20001, USA}\\
\footnotesize{$^{10}$ Imperial College London, South Kensington Campus, London SW7 2AZ, UK}\\
\footnotesize{$^{11}$Department of Physics and Astronomy, University of Iowa, IA 52242, USA}\\
\footnotesize{$^{12}$Institute for Space Astrophysics and Planetology,}\\
\footnotesize{Via del Fosso del Cavaliere 100, 00133 Rome, Italy }\\
\footnotesize{$^{13}$Space Sciences Laboratory,} \\ \footnotesize{University of California, Berkeley, CA 94720-7450, USA}\\
\footnotesize{$^\ast$Corresponding author; E-mail:  yeimy.rivera@cfa.harvard.edu.}
\footnotesize{$^\dagger$These authors contributed equally to this work.}
}

\date{}

\begin{document} 


\baselineskip24pt


\maketitle 


\section*{Abstract}
\begin{sciabstract}

After leaving the Sun's corona, the solar wind continues to accelerate and cools, but more slowly than expected for a freely expanding adiabatic gas. We use in situ measurements from the Parker Solar Probe and Solar Orbiter spacecrafts to investigate a stream of solar wind as it traverses the inner heliosphere. The observations show heating and acceleration of the the plasma between the outer edge of the corona and near the orbit of Venus, in connection to the presence of large amplitude Alfv\'en waves. Alfv\'en wave are perturbations in the interplanetary magnetic field that transport energy. Our calculations show the damping and mechanical work performed by the Alfv\'en waves is sufficient to power the heating and acceleration of the fast solar wind in the inner heliosphere.

\end{sciabstract}

In situ measurements have shown that the solar wind does not cool adiabatically as it expands away from the Sun \cite{Hellinger2011}. The speed and temperature profiles of the fast solar wind (at the highest speeds when measured far from the Sun) requires mechanical forcing and direct heating of the plasma after it leaves the solar atmosphere \cite{Parker1965, Alazraki1971, Halekas2023}. 

The cooling rate of solar wind protons depends on the speed of each wind stream \cite{Hellinger2011}. Protons in the slowest solar wind cool roughly adiabatically as they convect away from the corona, while protons in faster solar wind cool slower \cite{Shi2022}. The radial electron temperature profile exhibits similar behavior, though varies less with the stream speed \cite{Dakeyo2022}. Plasma that cools slower than adiabatic requires that additional heating occurs after the fast solar wind leaves the corona. The source of that heating is unknown.

Alfv\'en waves are transverse magnetohydrodynamic waves that travel along the magnetic field. Alfv\'en waves are thought to play a role in the processes that heat the solar wind \cite{Belcher1971, Alazraki1971, Chandran2009}. The energy budget of the solar wind indicates that energy provided by Alfv\'en waves makes a greater contribution to stream acceleration in higher solar wind speeds \cite{Halekas2023}. The acceleration of the solar wind streams with slower speeds is explained without the contribution from Alfv\'en waves \cite{Halekas2022}. Alfv\'en waves therefore more important to the dynamics of fast speed wind \cite{Reville2020, Shoda2021}.

In situ spacecraft measurements have found high-amplitude magnetic field rotations, termed switchbacks, near perihelion passes \cite{Bale2019, Kasper2019}. These switchbacks are characterized by a rapid change in the magnetic field direction with near constant magnetic field magnitude accompanied by correlated velocity fluctuations. Switchbacks have been interpreted as large amplitude Alfv\'en waves in the solar wind \cite{Kasper2019,Bale2019} however their definition and implied origin are debated. Although the switchback terminology suggests a change in polarity, the magnetic field does not always physically switchback to change its magnetic polarity. Groups of many such fluctuations have been found in coherent patches \cite{Bale2021}. The substantial wave energy associated with these large Alfv\'enic fluctuations close to the Sun, and their gradual evolution with heliocentric distance, indicates they could play a role in heating and acceleration of the solar wind \cite{Tenerani2021, Jagarlamudi2023}.

\textbf{Experimental design} A test of that hypothesis would be to measure the energy contributions in a switchback patch at points near and far from the Sun. Appropriate alignments with multiple spacecraft are rare \cite{Schwartz1983}, so statistical studies have combined measurements at similar heliocentric distances and with similar velocities \cite{Maksimovic2020, Dakeyo2022, Halekas2023}. Close to the Sun, spacecraft measurements show an overall decrease in speed of all solar wind. This indicates that acceleration is taking place and the statistical approaches taken breaks down. Alternative approaches separate the solar wind into percentiles \cite{Maksimovic2020, Dakeyo2022, Halekas2022} or by combining measurements taken at several heliocentric distances \cite{Halekas2023}, but these cannot isolate the evolution of individual plasma streams over large distances.

We investigate data from two spacecraft: Parker Solar Probe [\cite{Fox2016}, hereafter Parker] and Solar~Orbiter \cite{Muller2020}. Parker orbits close to the Sun, down to radii of 0.063 astronomical units (au), while Solar Orbiter is located further out with perihelia of 0.30 au. Our goal is to compare the properties and energetics of the same stream at different heliocentric distances, to determine its evolution as it moves outwards from the Sun. This requires identifying conjunctions when the spacecraft intersect the same solar wind stream. We expect the Parker data taken close to the Sun to show large amplitude Alfv\'en waves, while the Solar Orbiter data are sufficiently far out for the waves to have mostly dissipated.  

\paragraph*{Multi-spacecraft observations}

A suitable conjunction occurred in February 2022, when Parker and Solar~Orbiter crossed the same wind streamline at the same solar latitude within two days of each another. Parker crosses the stream when it was at 13.3 solar radii ($R_{\odot}$) from the Sun, the outer edge of the Alfv\'en region, defined as where the solar wind is slower than the local Alfv\'en wave speed \cite{Kasper2021}. The same plasma stream was subsequently crossed by Solar Orbiter at $127.7R_{\odot}$ where the solar wind is much faster than the Alfv\'en speed.

\begin{figure}[]
	\centering
	\includegraphics[width=0.9\linewidth]{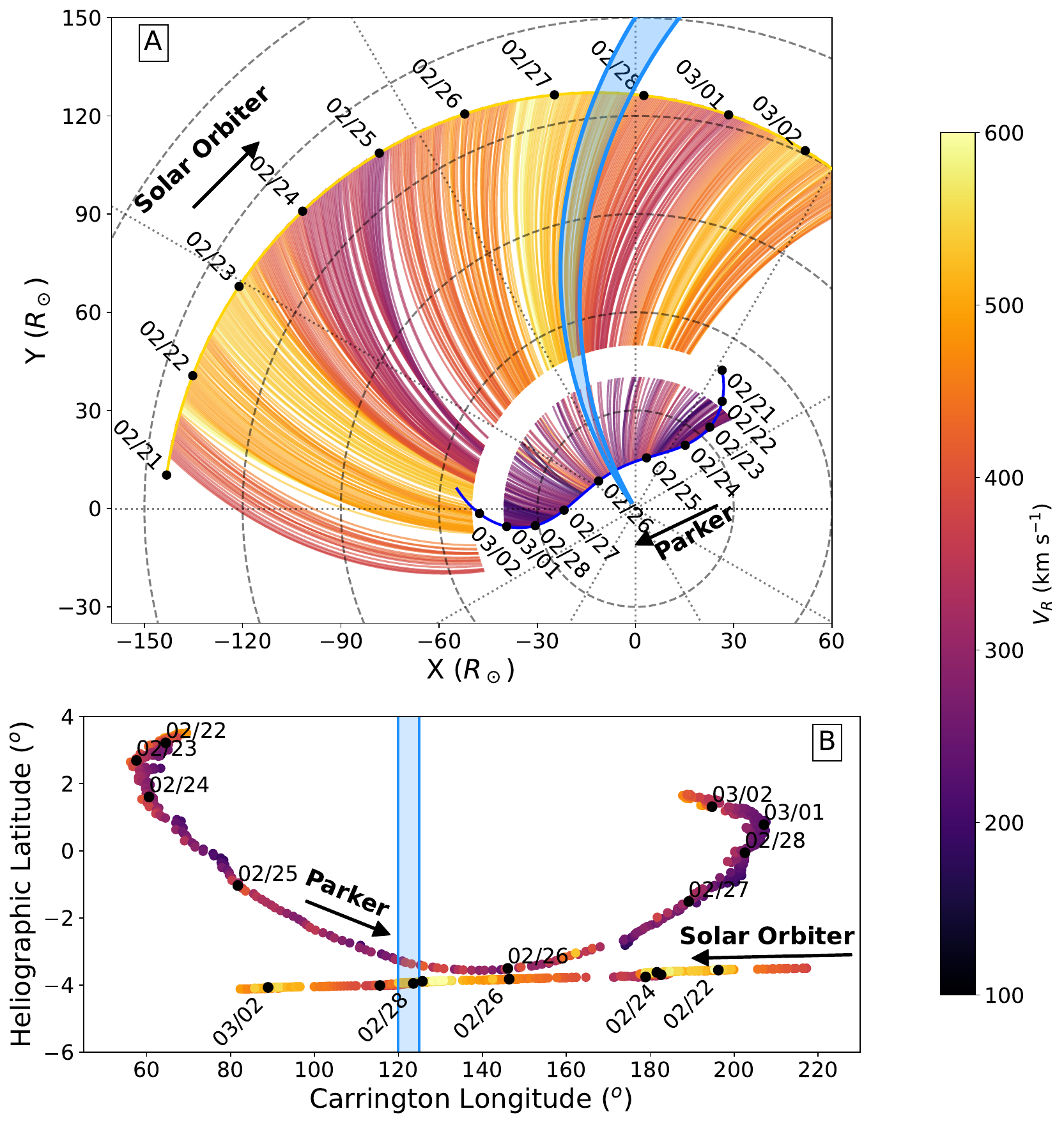}
	\caption{\footnotesize{\textbf{Spacecraft trajectories and ballistic mapping in the co-rotating reference frame.} (A) Solar~Orbiter and Parker trajectories projected onto the solar equatorial plane (Cartesian coordinates X and Y with the origin at the Sun). Spiral streamlines colored with the measured solar wind speed show the mapping of the spacecraft measurements to the Sun with a gap separating a segment generated from Solar Orbiter (outer segment) and Parker data (inner segment). Dotted and dashed lines show uniform spacing in longitude and radius respectively. Black arrows indicate the direction of spacecraft motion. The blue region indicates the fast solar wind stream we study. (B) The mapped heliographic coordinates of Parker and Solar~Orbiter at $2.5R_{\odot}$ colored as with the spirals in Panel (A). Parker's latitude is shifted up by 0.3$^{\circ}$ for visibility. The blue region, black circles and black arrows are as in Panel A.
 } }
    \label{fig:Backmapping}
\end{figure}

Figure~\ref{fig:Backmapping} shows this conjunction in the solar-corotating (Carrington) reference frame in panel A and across longitude and latitude in panel B. As done in \cite{Stansby2019, Badman2020}, we ballistically backmapped the stream trajectories back to 2.5$R_\odot$ above the Sun to determine the source surface longitude separately for the plasmas observed at Parker and Solar~Orbiter. The ballistic mapping uses the measured wind speeds and the respective spacecraft locations to define spiral trajectories (straight lines in the rotating reference frame) that approximately trace each stream back to its point of origin in the corona \cite{Nolte1973, Macneil2022, Dakeyo2024}. 

\begin{figure}[]
	\centering
	\includegraphics[width=0.98\linewidth]{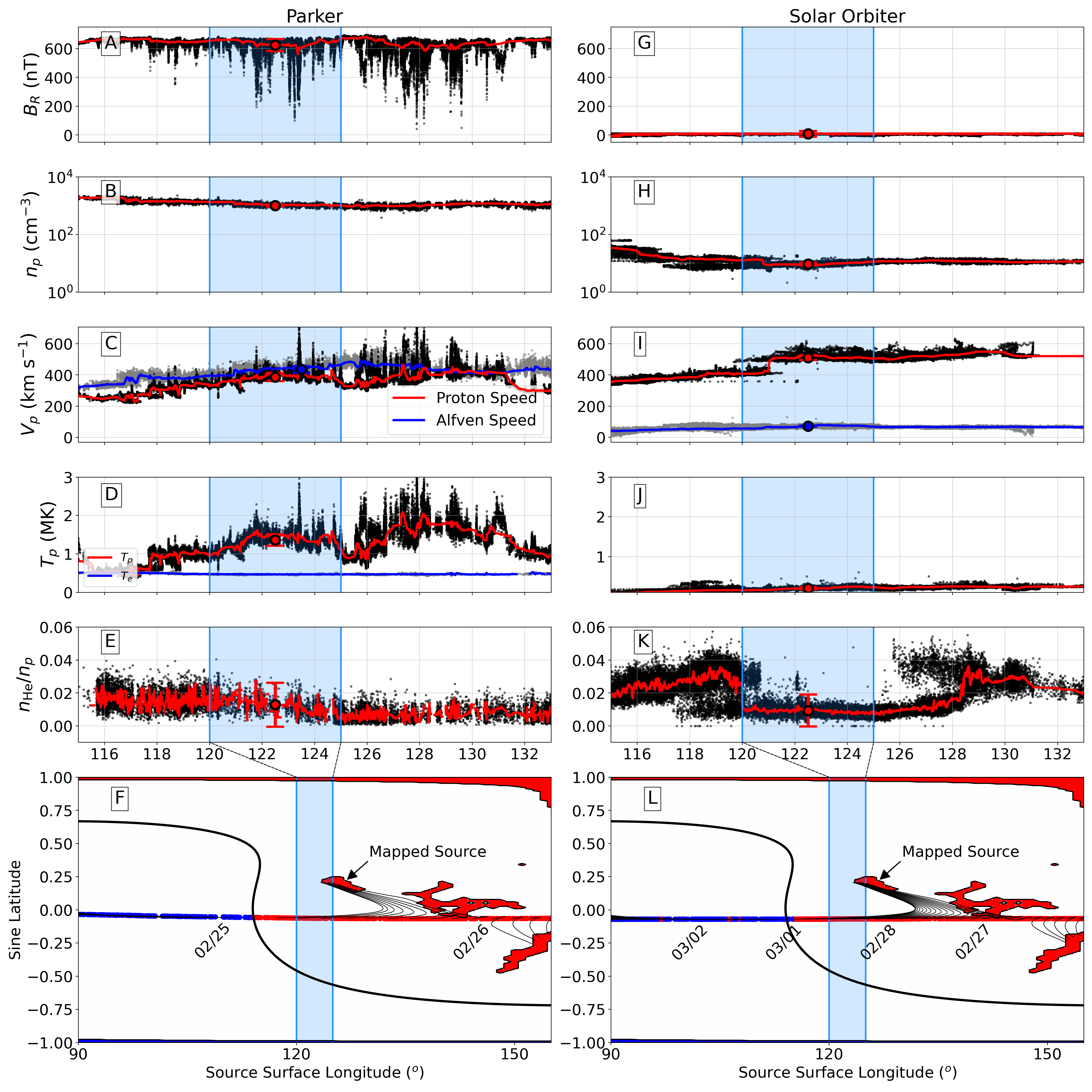}
	\caption{\footnotesize{\textbf{Solar wind properties measured by Parker (A-F) and Solar~Orbiter (G-L) across source surface longitude.} The blue shaded region indicates the fast solar wind stream we study. (A \& G) Radial magnetic field component, (B \& H) proton density, (C \& I) proton bulk speed and Alfv\'en velocity, (D \& J) isotropic proton temperature, (E \& K) n$_{\text{He}}$/n$_{p}$ density ratio. The black points show the measured values, while red solid curves are smoothed with a 50-point median filter. For panels where a second measurement is shown (C, D, I, J), the data and median filter are differentiated with grey points and a blue solid curve, respectively. Large circles with error bars (1 sigma) are values from Table~S1. (F \& L) Coronal hole source mapping. Colored points indicate the mapped positions where $B_R$ is positive (red) or negative (blue) and date labels as month/day. Black lines are field lines from a Potential Field Source Surface (PFSS) model linking the trajectory at the source surface to their solar sources at 1$R_\odot$, indicated by solid red and blue regions with black contours. The solid black curve shows the polarity inversion line. Both spacecraft map to the same location shown by the black arrows corresponding with in situ compositional context \cite{MaterialsandMethods}.}}
    \label{fig:Parker_solo_insitu_data}
\end{figure}

A stream of fast wind passes by both spacecraft during this conjunction (Fig.~\ref{fig:Parker_solo_insitu_data} and S1). The segment corresponds to source surface longitudes 120$^{\circ}$ to 125$^{\circ}$, in which the Parker data show a patch of large amplitude Alfv\'en waves which we term a switchback patch. Parker crossed this stream on 2022 February~25 (15:00--16:40 Coordinated Universal Time, UTC). Solar~Orbiter data show the same stream passing the spacecraft on 2022 February~27 (09:00--17:00 UTC). Parker crosses the same span of source longitude in a much shorter time because of its higher angular velocity, and the two spacecraft crossed the stream in opposite directions. The transit time for plasma to travel from the location of Parker to Solar~Orbiter during this period was 45~hours, estimated from our modeled velocity profile, similar to the delay between Parker exiting and Solar~Orbiter entering the stream ($\sim$~40~hours) \cite{MaterialsandMethods}. This implies that the plasmas encountered by Parker and Solar Orbiter were released from the Sun at roughly the same time, as well as from the same source.

The switchback patch corresponds to a local maximum in the solar wind speed appears in both Parker and Solar~Orbiter data taken over this longitude range. We identify further evidence of a connection between the plasma streams by identifying a positive polarity magnetic field at the same ballistically mapped longitude, 120--125$^o$, crossed by Parker and Solar Orbiter. We also find a consistent helium abundance around $1\pm0.5\%$ by number density (n$_\text{He}$/n$_\text{H} = 0.01$) in both solar wind streams. We expect plasma composition to remain fixed after leaving the corona. 

Over source surface longitudes 120--121$^{\circ}$, there is a bifurcation in the ballistic mapping of the Solar Orbiter data (Fig. 2K), showing a solar wind stream from the same source longitude that occurs later in the timeseries, with a He abundance of 3\%. Ballistic mapping does not account for stream interactions, allowing streams to unphysically overlap. We use the He abundance to disambiguate the fast stream from the slow stream, and exclude the latter from our analysis. The fast stream of interest maps back to a small equatorial coronal hole (Fig. \ref{fig:Parker_solo_insitu_data} F \& L). That point of origin is consistent with heavy ion abundances measured by Solar Orbiter, which are typical of coronal holes \cite{MaterialsandMethods}. Lastly, mass and magnetic flux density is conserved \cite{MaterialsandMethods}.

\paragraph*{Evolution of the plasma stream}
The plasma stream has different speeds at the two spacecraft, with the Parker data averaging $386\pm26$~km~s$^{-1}$ and the Solar~Orbiter data $512\pm15$~km~s$^{-1}$ (Table S1). We consider the later stream to be fast solar wind, even though the speed at Parker would be classified with slow solar wind if it were observed at Earth. The stream has undergone acceleration during its passage from $13.3R_\odot$ to $127.7R_\odot$ from the Sun. 

We compared the mass flux and magnetic flux at the two spacecraft to verify whether these quantities are conserved, as we expect for a expanding flux tube. We find that both quantities are conserved and that the stream compresses slightly more than the 1/$r^2$ variation we expect for flux tube (Fig. S2). We calculate a $10\pm9\%$ compression factor at Solar Orbiter (Eqn. S1, S2). For the energy flux budget of the stream, we find that energy is conserved within the measurement uncertainties, and the dominant source of uncertainty is the variance of the stream over the source surface longitude. The energy flux at Parker is $45.7\pm6.6$ and Solar Orbiter is $48.0\pm3.6$. We express the conservation of energy along the flux tube as:

\begin{align}
   W_{\text{Solar Orbiter}}-W_\text{Parker} = \Delta W &=
   \Delta \bigg( u_{\text{cm}}\Omega  \frac{r^2}{R_\odot^2} \bigg[U_{\text{K}} + U_{\text{H}} + U_{\text{G}}+ U_{\text{w}} \bigg]\bigg)    \label{eq:Energyconservation1}\\  
    &= \Delta W_{\text{K}} + \Delta W_{\text{H}} + \Delta W_{\text{G}} + 
     \Delta W_{\text{w}} \label{eq:Energyconservation2}\\ &= 0 \label{eq:Energyconservation3}
\end{align}

where the subscripts indicate kinetic (K), enthalpy (H), gravitation (G) and Alfv\'enic wave (w) energies. $\Delta$ denotes taking differences between quantities at Parker and Solar Orbiter.  The plasma center of mass velocity is computed as, $u_{\text{cm}}= \Sigma_j m_j n_j v_j/\Sigma_j m_j n_j$, where $j$ is each species: electrons, protons and alpha particles. $\Omega$ is the stream angular size in steradians at each spacecraft. Each energy density term ($U$) corresponds to a scaled energy flux term ($W$) once normalized by $u_{\text{cm}}$r$^2$/R$_{\odot}^2$ where r is the heliocentric distance \cite{Liu2021}. Each $U$ and $W$ term includes contributions from electrons, protons and alpha particles. This formulation does not explicitly include the ambipolar electric potential induced by hot electrons escaping the Sun's corona; that energy is included in the electron thermal pressure or the enthalpy ($U_{\text{H}}$) \cite{Halekas2022, Halekas2023}. Equations \ref{eq:Energyconservation1}-\ref{eq:Energyconservation3} indicate the total energy transported through a cross sectional area $\Omega r^2$ at center of mass velocity $u_{cm}$ is equal at both spacecraft. We compute the contributions of each term from the Parker and Solar Orbiter measurements \cite{MaterialsandMethods}. We take $\Omega = 1$ at Parker and compute an equivalent expansion factor ($f$) at Solar Orbiter through consideration of mass flux conservation \cite{MaterialsandMethods}.

\begin{figure}[]
	\centering
	\includegraphics[width=\linewidth]{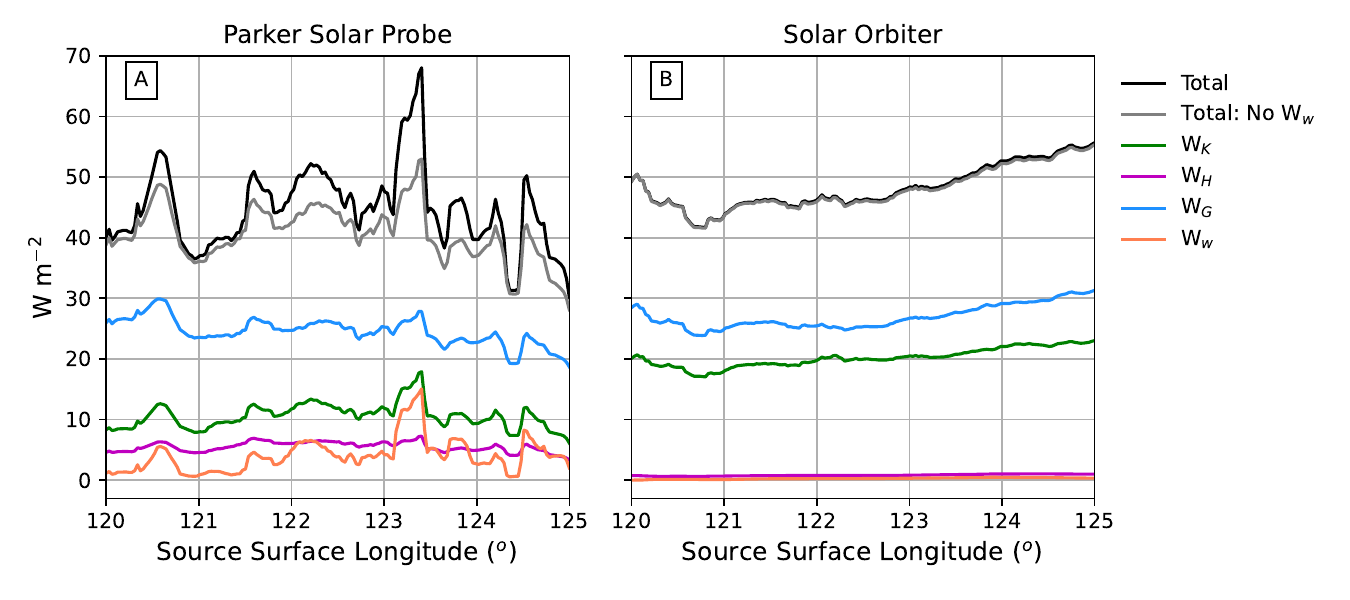}
	\caption{\textbf{Energy fluxes across the fast wind stream.} Panels show the individual energy flux terms for Parker (A) and Solar~Orbiter (B, multiplied by the expansion factor, $f$ \cite{MaterialsandMethods}). The grey/black comparison indicates the switchback wave energy at Parker is needed for energy conservation.}
	\label{fig:Parker_solo_energy}
\end{figure}

The resulting energy flux terms are shown as a function of source surface longitude for Parker (Fig. \ref{fig:Parker_solo_energy}A) and Solar~Orbiter (Fig. \ref{fig:Parker_solo_energy}B). We find that energy conservation is satisfied to within the standard deviation of the total energy flux, but only when wave energy is included. Uncertainties are computed as the standard deviation of the energy flux terms across the 5 degree source surface longitude range. The instrumental uncertainties are smaller than the intrinsic variability of the measurements. The overall wave energy term is of similar magnitude to the total energy flux variability, but contributes a substantial part of the energy budget at Parker and its inclusion is required to maintain energy conservation between the locations of the two spacecraft (Table~S2). 

\begin{figure*}[]
	\centering
	\includegraphics[width=\linewidth]{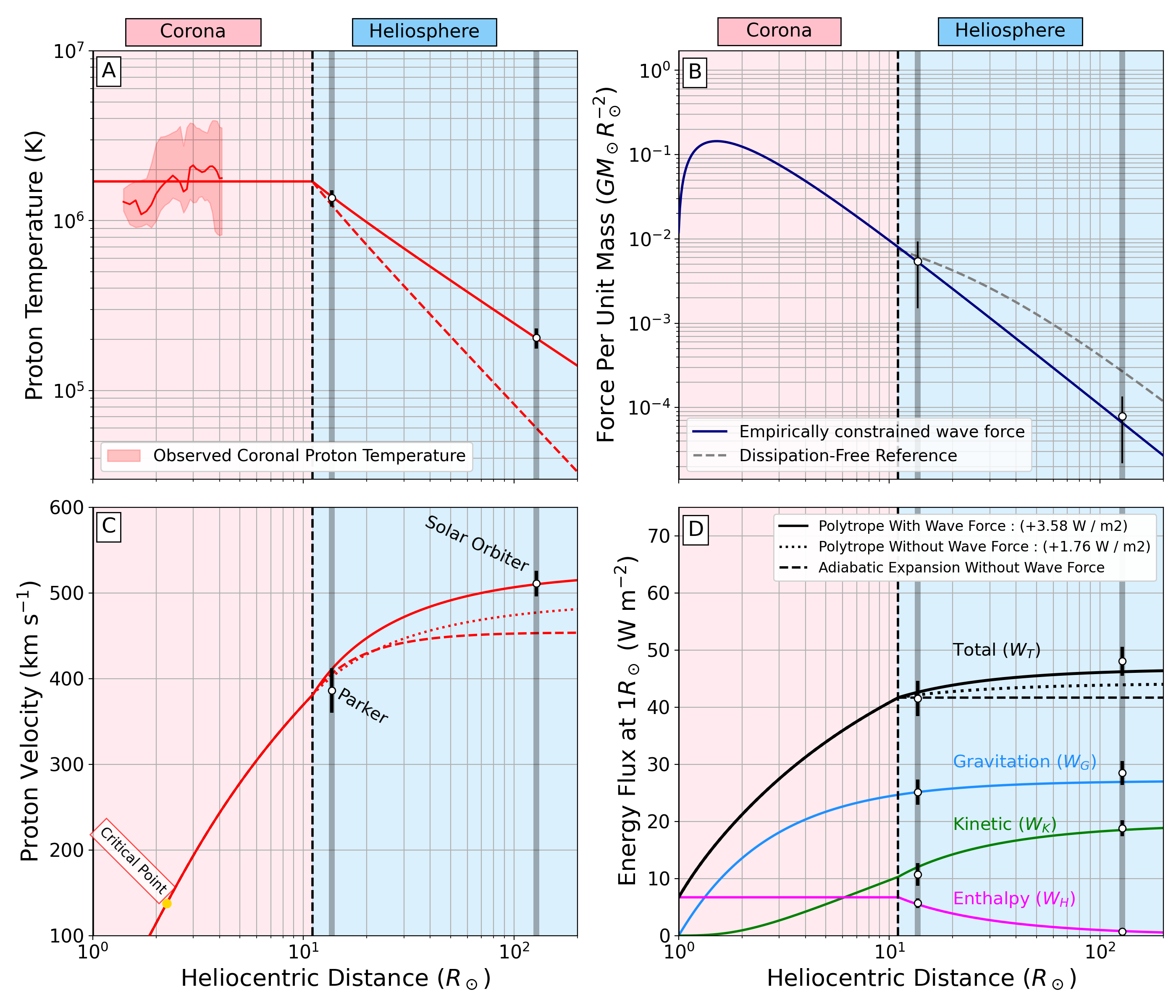}
	\caption{ \footnotesize{\textbf{Comparison of measured properties with a solar wind model across heliocentric distance.} The white datapoints and 1 sigma errorbars are measurements at Parker and Solar Orbiter reported in Table S1 and S2, and curves are model results. Modeling is split into a coronal and heliospheric portion (pink and blue background shading, respectively). (A) Proton temperature for polytropic (solid) and adiabatic (dashed) expansion.  Remote observations of fast wind low in the corona \cite{Cranmer2020a} in red shading and an associated red line. (B) Wave pressure gradient force using a data-constrained analytical model (solid blue, \cite{MaterialsandMethods}) and a dissipation-free curve (dashed grey, \cite{MaterialsandMethods}). (C) Proton velocity for adiabatic expansion only (dashed red), polytropic expansion only (red dotted), polytropic expansion and wave pressure (solid red). The critical point is in yellow. (D) Energy flux terms: Enthalpy (magenta), kinetic (green), gravitation (blue) and the total (black). The total is shown for each case as in (C).}}
 
	\label{fig:polytropic_model}
\end{figure*}

\paragraph*{Acceleration and heating of the stream}
If we assume that the stream did not change in time over this period and the plasma obeys a polytropic equation of state, then the expansion profile for the stream of solar wind is fully determined by hydrodynamics \cite{Parker1958, Parker1960}. Following \cite{Parker1958, Dakeyo2022, Shi2022}, we use a polytrope function, $P\rho^{-\gamma}=C$, to extrapolate from the Parker and Solar Orbiter observations, where $P$ is the plasma thermal pressure, $\rho$ the mass density, $\gamma$ is the polytropic index, and $C$ is a constant. Figure ~\ref{fig:polytropic_model} shows the resulting proton temperature, wave pressure force, and proton speed profiles, and energy flux evolution of the stream. Fig. \ref{fig:polytropic_model}C \& D compare three choices of polytrope (i) free adiabatic expansion, $\gamma = 5/3$, (ii) a polytrope in which the temperature was derived from a model is fitted to the observations ($\gamma = 1.41 \pm 0.020$), and (iii) the same fitted polytrope with the addition of empirically-constrained wave pressure force profile \cite{MaterialsandMethods}.  In the latter case, the mechanical work performed by the Alfv\'en waves was determined from an analytical force profile \cite{Shi2022} (Fig. \ref{fig:polytropic_model}B), which is consistent with the measured gradient in Alfv\'enic wave pressure \cite{MaterialsandMethods}. 

Figure ~\ref{fig:polytropic_model} also shows extrapolations of these polytrope models into the corona, where we assume the temperature profile (Fig. \ref{fig:polytropic_model}A) is approximated as constant, following previous work \cite{Parker1958}. They consist of a constant proton temperature, $T_{iso}=1.7\times10^6$ K (Fig. \ref{fig:polytropic_model}A), and combined with the wave pressure force profile (Fig. \ref{fig:polytropic_model}B) within a radius of $R_{iso} = 11~R_\odot$, are defined out from the center of the Sun. These parameters were chosen based on remote measurements of the fast solar wind temperature in the corona measured from ultraviolet remote observations \cite{Cranmer2020a}, and are consistent with other remote observations that showed the sonic critical point, where the solar wind speed exceeds the local sound speed, is located at 1.9~$R_\odot$\cite{Telloni2019} near our modeled value of 2.2$R\odot$ (Fig. \ref{fig:polytropic_model}C). We extrapolate that the coronal wave pressure forcing (Fig. \ref{fig:polytropic_model}B) contributes  about 20 W m$^{-2}$ to the energy budget compared to 120 W $m^{-2}$ which is present at the base of the corona \cite{DePontieu2007}. This extrapolation to the corona indicates that our results are consistent with the expected physical parameters of the corona. Under the isothermal conditions we assume at the corona, the plasma still receives energy from the Alfv\'en waves, that we assumed to exactly balance the cooling due to expansion.

Figure~\ref{fig:polytropic_model}C, D shows the effect of the Alfv\'en wave energy flux on the system. All three polytrope models shown start from the same initial conditions at R$_{iso}$, but only the polytrope that includes both Alfv\'en wave pressure forcing and heating matches the measured acceleration between Parker and Solar~Orbiter (Fig.~\ref{fig:polytropic_model}C). We estimate the mechanical work done from the measured decline in wave pressure ($1.88\pm0.31$~W~m$^{-2}$) and from the shallower-than-adiabatic thermal pressure gradient of the polytropic heating profile ($1.76\pm0.25$~W~m$^{-2}$). The portion of the wave energy flux resulting in acceleration ($1.88\pm0.31$~W~m$^{-2}$) is substantially less than the mean of the total wave energy flux lost ($3.9\pm2.7$~W~m$^{-2}$). This implies that the large amplitude Alfv\'en waves are damped. If there was no damping, all the wave energy flux would be converted to mechanical work, and the wave force would be shallower and follow the dissipation free curve (Fig.~\ref{fig:polytropic_model}B) as discussed in \cite{Belcher1971,Alazraki1971,Jacques1977}. The unused wave energy flux is similar to the energy input required to sustain the temperature profile with our fitted polytropic index. We speculate that the waves are converted to heat via reflection driven turbulence \cite{Chandran2009, vanBallegooijen2016}, and turbulent dissipation \cite{Sorriso-Valvo2007, MacBride2008, Stawarz2009}.

Observations directly show that substantial heating and acceleration of the plasma, above that expected for free adiabatic expansion ($3.62\pm0.40$~W~m$^{-2}$) occurred in this region. The large amplitude Alfv\'en waves organized in coherent patches dissipate $3.9\pm2.9$~W~m$^{-2}$ along the stream. These Alfv\'enic structures can therefore provide the necessary additional heating and acceleration as the solar wind moves through the corona and inner heliosphere.

%%%%%%%%%%%%%%%%%%% FIG 1 %%%%%%%%%%%%%

%%%%%%%%%%%%%%%%%%%%%%%%%%%%%%%%%%%%%%%

%%%%%%%%%%%%%%%%%%% FIG 2 %%%%%%%%%%%%%

%%%%%%%%%%%%%%%%%%%%%%%%%%%%%%%%%%%%%%%

%%%%%%%%%%%%%%%%%%% FIG 3 %%%%%%%%%%%%%

%%%%%%%%%%%%%%%%%%%%%%%%%%%%%%%%%%%%%%%

%%%%%%%%%%%%%%%%%%% FIG 4 %%%%%%%%%%%%%

\section*{Acknowledgements}

Parker Solar Probe was designed, built, and is now operated by the Johns Hopkins Applied Physics Laboratory as part of NASA’s Living with a Star (LWS) program (contract NNN06AA01C).  The SWEAP Investigation is a multi-institution project led by the Smithsonian Astrophysical Observatory in Cambridge, Massachusetts. Other members of the SWEAP team come from the University of Michigan, University of California, Berkeley Space Sciences Laboratory, The NASA Marshall Space Flight Center, The University of Alabama Huntsville, the Massachusetts Institute of Technology, Los Alamos National Laboratory, Draper Laboratory, JHU’s Applied Physics Laboratory, and NASA Goddard Space Flight Center. The FIELDS instrument suite was designed and built and is operated by a consortium of institutions including the University of California, Berkeley, University of Minnesota, University of Colorado, Boulder, NASA/GSFC, CNRS/LPC2E, University of New Hampshire, University of Maryland, UCLA, IFRU, Observatoire de Meudon, Imperial College, London and Queen Mary University London.

Solar Orbiter is a space mission of international collaboration between ESA and NASA, operated by ESA.  The Solar Orbiter Solar Wind Analyser (SWA) data are derived from scientific sensors which have been designed and created, and are operated under funding provided in numerous contracts from the UK Space Agency (UKSA), the UK Science and Technology Facilities Council (STFC), the Agenzia Spaziale Italiana (ASI), the Centre National d’Etudes Spatiales (CNES, France), the Centre National de la Recherche Scientifique (CNRS, France), the Czech contribution to the ESA PRODEX programme and NASA. The Solar Orbiter magnetometer was funded by the UK
Space Agency (grant ST/T001062/1).

\textbf{Funding:} Y.J.R. acknowledges support from the Future Faculty Leaders postdoctoral fellowship at Harvard University and NASA grant NNN06AA01C.. 
S.T.B., M.L.S., K.W.P., T.N., S.D.B, J.S.H., acknowledges support from NASA grant NNN06AA01C.
R.M.D. acknowledges support from NASA grant 80NSSC22K0204. 
J.L.V. acknowledges support from NASA PSP-GI grant 80NSSC23K0208 and NASA LWS grant 80NSSC22K1014. 
J.E.S. is supported by the Royal Society University Research Fellowship URF\textbackslash R1\textbackslash 201286. 
S.A.L. was provided by NASA contract NNG10EK25C. 
S.T.L., J.M.R. was provided from the University of Michigan was provided through SwRI subcontract A99201MO. 
T. S. H. is supported by STFC grant ST/W001071/1.
C.J.O., P.L., R.D.M. is supported by STFC grants ST/T001356/1 and ST/S000240/1.
J.S.H. is supported by NASA grant 80NSSC22K1014.

\textbf{Authors contributions:} Y.J.R. and S.T.B. were jointly responsible for the conceptualization, formal analysis, and writing the manuscript. M.L.S, K.W.P., T.N., C.S., T.S.H. helped conceive and design the study. J.E.S., R.D.M., J.S.H contributed interpretation of the spacecraft data. J.L.V., R.M.D., S.A.L., P.L., C.J.O., J.C.K., T.S.H., S.D.B., J.M.R., S.T.L. performed data acquisition. All authors commented on the manuscript. 

\textbf{Competing interests:} We have no conflicts of interest to report.

\textbf{Data and materials availability:} For Parker Solar Probe the SWEAP data are available at, \url{http://sweap.cfa.harvard.edu/pub/data/sci/sweap/spi/L2/spi_sf00/2022/02/} and the FIELDS data at \url{https://research.ssl.berkeley.edu/data/psp/data/sci/fields/l2/mag_RTN_4_Sa_per_Cyc/2022/02/}. All Parker Solar Probe data are in the CDF files for 2022 February 25. The Solar Orbiter data are available at \url{https://soar.esac.esa.int/soar/#search}. All Solar Orbiter data are from CDF files from 2022 February 27 from instruments MAG and SWA, with a processing level 2. Our modeling code is available at \url{https://github.com/STBadman/ParkerSolarWind}  and has been archived at Zenodo \cite{Badman2023_software}. Our measured properties of the stream are listed in Table S1, and the results of our energy budget calculations in Table S2.

\section*{Supplementary Materials}
\begin{itemize}
    \item Materials and Methods
    \item Supplementary Text
    \item Figs.~S1~to~S4, Table~S1, S2
    \item References \cite{Kasper2016, Livi2022, Whittlesey2020, Halekas2020, Bale2016, Horbury2020, Livi2023, Owen2020, Demarco2023,  Romeo2023, Badman2023, Badman2023_software, Asplund2021, vonsteiger2000, Kasper2007, Woolley2021, McManus2022, Fisk2020, Zank2020, Bale2023}
\end{itemize}

\bibliographystyle{Science}

\end{document}